\documentclass[12pt]{aastex631}
\usepackage{color}
\usepackage{hyperref}
\usepackage{epstopdf}
\usepackage{graphicx}
\usepackage[FIGTOPCAP]{subfigure}
\usepackage{CJKutf8}
\usepackage{amsmath}

\begin{document}
	%\linenumbers 
	\title{GRB 200716C: Evidence for a Short Burst Being Lensed}
	
	\correspondingauthor{Yi-Zhong Fan}
	\email{yzfan@pmo.ac.cn}
	
	\author{Yun Wang}
	\affiliation{Key Laboratory of Dark Matter and Space Astronomy, Purple Mountain Observatory, Chinese Academy of Sciences, Nanjing 210034, China}
	\affiliation{School of Astronomy and Space Science, University of Science and Technology of China, Hefei, Anhui 230026, China}
	
	\author{Lu-Yao Jiang}
	\affiliation{Key Laboratory of Dark Matter and Space Astronomy, Purple Mountain Observatory, Chinese Academy of Sciences, Nanjing 210034, China}
	\affiliation{School of Astronomy and Space Science, University of Science and Technology of China, Hefei, Anhui 230026, China}
	
	\author{Cheng-Kui Li}
	\affiliation{Key Laboratory of Particle Astrophysics, Institute of High Energy Physics, Chinese Academy of Sciences, 19B Yuquan Road, Beijing 100049, People’s Republic of China}
	
	\author{Jia Ren}
	\affiliation{School of Astronomy and Space Science, Nanjing University, Nanjing 210093, China}
	\affiliation{Key Laboratory of Modern Astronomy and Astrophysics (Nanjing University), Ministry of Education, China}
	
	\author{Shao-Peng Tang}
	\affiliation{Key Laboratory of Dark Matter and Space Astronomy, Purple Mountain Observatory, Chinese Academy of Sciences, Nanjing 210034, China}
	\affiliation{School of Astronomy and Space Science, University of Science and Technology of China, Hefei, Anhui 230026, China}
	
	\author{Zi-Min Zhou}
	\affiliation{Guangxi Key Laboratory for Relativistic Astrophysics, School of Physics Science and Technology, Guangxi University, Nanning 530004, China}
	
	\author{Yun-Feng Liang}
	\affiliation{Guangxi Key Laboratory for Relativistic Astrophysics, School of Physics Science and Technology, Guangxi University, Nanning 530004, China}
	
	\author{Yi-Zhong Fan}
	\affiliation{Key Laboratory of Dark Matter and Space Astronomy, Purple Mountain Observatory, Chinese Academy of Sciences, Nanjing 210034, China}
	\affiliation{School of Astronomy and Space Science, University of Science and Technology of China, Hefei, Anhui 230026, China}
	\begin{abstract}
		%The radiation of a tiny fraction of gamma-ray bursts (GRBs) may be lensed before reaching the earth.
		A tiny fraction of observed gamma-ray bursts (GRBs) may be lensed.
		The time delays induced by the gravitational lensing are milliseconds to seconds if the point lenses are intermediate-mass black holes.
		The prompt emission of the lensed GRBs, in principle, should have repeated pulses with identical light curves and spectra but different fluxes and slightly offset positions.
		In this work, we search for such candidates within the GRBs detected by Fermi/GBM, Swift/BAT, and HXMT/HE and report the identification of an attractive event GRB 200716C that consists of two pulses.  
		Both the autocorrelation analysis and the Bayesian inference of the prompt emission light curve are in favor of the gravitational lensing scenario.
		Moreover, the spectral properties of the two pulses are rather similar and follow the so-called Amati relation of short GRBs rather than long duration bursts.
		The measured flux ratios between the two pulses are nearly  constant in all channels,  as expected from gravitational lensing.
		We therefore suggest that the long duration burst GRB 200716C was a short event being lensed. The redshifted mass of the lens was estimated to be $4.25^{+2.46}_{-1.36}$ $\times$ $10^5$ $M_{\odot}$ (90$\%$ credibility).
		If correct, this could point towards the existence of an intermediate-mass black hole along the line of sight of GRB 200716C.
	\end{abstract}
	
	
	\section{introduction}
	Gravitational lensing takes place when the lights coming from a distant source are bent by a clump of massive matter (such as globular cluster, dark matter halo, and black hole) as they travel towards the observer.
	For a strong gravitational lensing, there are visible distortions such as the formation of Einstein rings, arcs, and multiple images.
	GRBs occured at cosmological distances.  The radiation of a tiny fraction of GRBs could be strongly lensed by the foreground objects before reaching us  \citep{paczynski1986gamma,paczynski1987gravitational,mao1992gravitational}. 
	Lensed GRBs should have two groups of repeated pulses with quasi identical shape and spectrum but different fluxes and slightly offset locations. 
	Due to the poor angular resolution of the gamma-ray instruments, however, the  small location offsets are not possible to be reliably resolved.  
	Fortunately, the instruments usually have a high time resolution and the gravitational lensing-induced time delay is  detectable. 
	
	\par
	In 1990s, several methods had been proposed  to identify the lensed GRBs \citep{wambsganss1993method,nowak1994can} and since then dedicated efforts have been made to search for such events in the GRB data. 
	%Based on observations from different satellites, series of studies have been performed to search for the time delay in GRBs.
	For instance, \cite{li2014search}  studied 2,700 bursts observed by the BATSE.
	\cite{hurley2019search}  analyzed a sample of 2,301 GRBs detected by Konus-Wind.
	In the last decade,  the Fermi-GBM  data had been widely used to hunt for the lensed events \citep{veres2009search,2011AIPC.1358...17D,ahlgren2020search}.
	Anyhow, just null results have been reported in these works. 
	Very recently, \cite{paynter2021evidence} have developed a Bayesian inference-based method to identify gravitational lensing event and reported tentative evidence for a lensed GRB.
	These previous approaches can be classified into two groups. One is to analyze the autocorrelation between pulses in the same GRB and the other focuses on the cross-correlation between different GRBs. In this work we concentrate on the former and search for the lensing bursts with a large GRB database, including 3,099 Fermi/GBM events, 1,297 Swift/BAT events, and 311 HXMT/HE events.
	GRB 200716C, one burst detected simultaneously by all these three instruments, is found to be a promising candidate.

	\par
	This paper is organized as follows.
	In Section 2, we introduce the observations of GRB 200716C and perform the autocorrelation analysis of the light curves in different energy bands.  
	In Section 3, we provide further evidence for GRB 200716C as a lensing event.
	In Section 4, we summarize our results with some discussion.
	
	\section{Observation and the autocorrelation analysis }
	\subsection{GRB 200716C}
	The prompt emission of  GRB 200716C was observed by multiple satellites.
	The Fermi GBM team reported the detection of a possible long GRB (trigger 616633066.180458/200716957) at 22:57:41 UT on 16 Jul 2020 \citep{2020GCN.28135....1V}.
	The GBM light curve consists of two separated pulses with a total duration \emph{T$_{\rm 90, GBM}$} $\sim5.3$ s in the $50-300$ keV band.
	At the same time, the {Swift} Burst Alert Telescope (BAT) triggered and located GRB 200716C (trigger=982707), and the emission in the $50-300$ keV band lasted for ${T_{\rm 90,BAT}}=5.44$ s \citep{2020GCN.28136....1B}.
	In addition, {HXMT/HE} detected GRB 200716C (trigger ID: HEB200716956) in a routine search of the data \citep[{HXMT/HE;}][]{2020GCN.28145....1X}.
	The {HXMT/HE} light curve has a duration of $T_{\rm 90,HXMT}\sim 2.16$ s.
	
	\par
	In this work, we adopt the time-tagged event
	(TTE) data of these three satellites.
	The Fermi/GBM data are provided by the public science support center (FSSC) of the Fermi satellite (http://fermi.gsfc.nasa.gov/ssc/data/). 
	The {Swift}/BAT data are available at the website (http://www.swift.ac.uk).
	The {HXMT/HE} data can be conveniently applied online (http://www.hxmt.org). Fermi/GBM has 12 sodium iodide (NaI) detectors and 2 bismuth germanate (BGO) detectors. 
	According to the angle between each detector and the source, we use the data of two NAI detectors (n0 and n1; $8-900$ keV) and one BGO detector (b0; $200-40000$ keV).
	The energy range of {Swift}/BAT is $15-350$ keV. While for {HXMT/HE}, the deposit energy range is $100-600$ keV in normal mode.
	\par
	We analyze the Swift/BAT data using standard {\tt HEASoft} tools (version 6.28).
	The processing of the Fermi/GBM data is with the {\tt GBM Data Tools} (https://fermi.gsfc.nasa.gov/ssc/data/analysis/gbm/), which is very convenient for user customization.
	For the HXMT/HE data, we use {\tt Astropy} ({https://www.astropy.org/index.html}) to manipulate the event file, and extract the light curve in different filters.
	We take the Fermi trigger time as the zero point and plot the light curve for each detector in different energy bands (see the left upper panel of Figure 1).
	%%%%%%%%%%%%%%%%%%%%%%%%%%%%%%%%%%%%%%%%%%%%%%%%%%%%%
	\subsection{Candidate check}
	Signal autocorrelation can be used to measure the time delay of temporally overlapping signals of a gravitationally lensed system.
	The standard autocorrelation function (ACF) is defined as
	\begin{equation}
		C(k) = \frac{\sum_{t=0}^{N-k} (I_t-\overline I)(I_{t+k}-\overline I)}
		{\sum_{t=0}^{N} (I_t-\overline I)^2}.
		\label{eq:correlation}
	\end{equation}
	\par
	We adopt the Savitzky-Golay filter $F(\delta t)$ to fit the ACF sequence. 
	The values of the window length and the order of the polynomial are set to be 101 and 3, respectively.
	The dispersion ($\sigma$) between the ACF and the fit $F(k)$ is
	\begin{equation}
		\sigma^2 = \frac{1}{N}\sum_{j=0}^{N} [C(k) - F(k)]^2,
		\label{eq:dispersion}
	\end{equation}
	where $N$ is the total number of bins.  As usual, 
	we identify the $3\sigma$ outliers as gravitational-lensing candidates.	
	\par
	We calculate the autocorrelation values of the light curves of GRB 200716C for different detectors and in multi energy channels, and perform the Savitzky-Golay filtering. The results are shown in the right panel of Figure 1.
	In all cases, the autocorrelation analyses yield significance greater than 3 $\sigma$,  indicating that the two pulses of GRB 200716C have rather similar temporal behaviors in each channel. The maximum significance ($5.08\sigma$) holds for the Fermi/GBM-NAI ($276-900$ keV) data. Furthermore, the delay times are nearly the same ($\approx 1.9$ s).  For point sources the gravitational lensing is achromatic, thus we expect that each energy channel of a lensed GRB should autocorrelate with the same time delay.  Motivated by these facts, we suggest that GRB 200716C is a candidate lensing event.  
	
	%%%%%%%%%%%%%%%%%%%%%%%%%%%%%%%%%%%%%%%%%%%%%%%%%%%%%
	\section{Lensing Analysis}
	In a gravitational lensing system, photons that travel longer distances arrive first, because a shorter path means that the light passes through deeper gravitational potential well of the lens, where the time dilation is stronger.
	The source flux is loweer for the photons coming relatively later than for those earlier.
	Consequently, for a lensed GRB there will be at least one early pulse followed by a weaker pulse.
	The time delay between these two pulses is determined by the mass of the gravitational lens.
	For lensing of a point mass, we have \citep{krauss1991new,narayan1992determination,mao1992gravitational}
	\begin{equation}
		(1+z_\text{l})M_l = \frac{c^3\Delta t}{2G}\left(\frac{r-1}{\sqrt{r}} +\ln r\right)^{-1}.
		\label{mass_redshift}
	\end{equation}
	where $\Delta t$ is the time delay, $r$ is the ratio of the fluxes of the two pulses, and $(1+z_\text{l})M_l$ is the redshifted lens mass.
	With the measured $\Delta t$ and $r$, it is straightforward to calculate the redshifted mass $(1+z_\text{l})M_l$.
	
	\subsection{Bayesian Inference}
	In order to clarify whether the two-pulse light curve is due to the lensing effect of a single pulse,
	\cite{paynter2021evidence} developed a Python package called {\tt PyGRB} (https://github.com/JamesPaynter/PyGRB) to create light-curves from either pre-binned data or time-tagged photon-event data.
	The {\tt PyGRB} was developed for the BATSE bursts.
	In this work we extend it to accommodate different gamma ray instruments such as Fermi/GBM, Swift/BAT and HXMT/HE.
	We use the Bayesian statistical framework to obtain the posterior distributions of the parameters.
	Bayesian evidence ($\mathcal{Z}$) is calculated for model selection and can be expressed as
	\begin{equation}
		\mathcal{Z} = \int \mathcal{L}(d|\theta) \pi(\theta) d\theta, 
	\end{equation}
	where $\theta$ is the model parameters, and $\pi(\theta)$ is the prior probability.
	For TTE  data from various instruments, the photon counting obeys a Poisson process and
	the likelihood $\ln\mathcal{L}$ for Bayesian inference takes the form of
	\begin{align}
		\ln {\cal L}(\vec{N}|\theta) = & \sum_i
		\ln{\cal L}(N_i|\theta) \\
		= & \sum_i N_i\ln\Big(\delta t_i B +  \delta t_i S(t_i|\theta)\Big) \nonumber\\
		& -  \Big(\delta t_i B +  \delta t_i S(t_i|\theta)\Big) -\log(N_i!),
	\end{align}
	where $N_i$ stands for observed photon count in each time bin, and the model predicted photon count  consists of the background count $\delta t_i B$ and the signal count $\delta t_i S(t_i|\theta)$.
	Note that the differences of $\mathcal{Z}$ among models are important for our purpose.  
	Hence we define different signal models $S(t_i|\theta)$ to describe whether the pulses are lensed images or not.
	Several functions have been proposed to describe the pulse shapes \citep{krauss1991new,narayan1992determination,mao1992gravitational,paynter2021evidence}. Here we adopt the fast-rising exponential decay (FRED) pulse light curve model 
	\begin{equation}
		S(t|\Delta,A,\tau,\xi) = A \exp \left[ - \xi \left(  \frac{t - \Delta}{\tau} + \frac{\tau}{t-\Delta}  \right)   \right],
	\end{equation}
	where $\Delta$ is the start time of pulse, $A$ is the amplitude factor, $\tau$ is the duration parameter of pulse, and $\xi$ is the asymmetry parameter used to adjust the skewness of the pulse. 
	In addition to the FRED pulses, the GRB light curve may be accompanied by a slow component \citep{vetere2006slow}, which is described by a  
	Gaussian function
	\begin{equation}
		S_{\rm gaus}\text(t|\Delta,A,\sigma)= A \exp \left[ -\frac{(t - \Delta)^2}{2\sigma^2} \right].
	\end{equation}
	With the above formulae, for the double pulse case (i.e., GRB 200716C) we  describe the lensing and null scenarios as  
	\begin{equation}
		S_\text{lens}(t|\theta_\text{lens}) = S(t|\Delta,A,\tau,\xi) + r^{-1} \cdot S(t|\Delta+\Delta_t,A,\tau,\xi) + B,
	\end{equation}
	\begin{align}
		S_\text{non-lens}(t|\theta_\text{non-lens}) =& S(t|\Delta_1,A_1,\tau_1,\xi_1) \nonumber\\
		&+ S(t|(\Delta_1+\Delta_t,A_2,\tau_2,\xi_2)  + B.
	\end{align}
	For lensing model, $r$ is the flux ratio between two pulses (see Eq. (3)) and \emph{B} is a constant background parameter.
	After adding a slow component, we have four models for Bayesian inference, including $S_\text{lens}$, $S_\text{lens,gaus}$, $S_\text{non-lens}$, and $S_\text{non-lens,gaus}$.
	%Bayesian inference will suggest whether the slow component is needed or not.
	$S_\text{non-lens}$ has three more parameters than $S_\text{lens}$. The influence of the number of parameters on the preference of the model, however, has been properly addressed in the calculation of $\mathcal{Z}$.
	We use the nested sampling algorithm Dynesty \citep{speagle2020dynesty,skilling2004nested,skilling2006nested,higson2019dynamic} in {\tt Bilby} \citep{ashton2019bilby} to sample the posterior distributions of all those parameters, typically with 500 live points.
	The ratio of the $\mathcal{Z}$ for two different models is called as the Bayes factor (BF)
	and the logarithm of the Bayes factor reads
	\begin{align}
		\ln(\text{BF}) = & \max(\ln{\cal Z}_\text{lens},\ln{\cal Z}_\text{lens,gaus}) \nonumber\\ 
		&- \max(\ln{\cal Z}_\text{non-lens},\ln{\cal Z}_\text{non-lens,gaus}) ,
		\label{com_bayes}
	\end{align}
	where the symbol ``$\max$" means taking the larger number between them.
	As a statistically rigorous measure for model selection, 
	if $\ln{\rm(BF)} > 8$ we have the ``strong evidence'' in favor of one hypothesis over the other \citep{thrane2019introduction}.
	The results of Bayesian inference for each model with different data are summarized in Table 1.
	\par
	{ To estimate the global significance of the candidate lensing event, following \cite{paynter2021evidence} we combine the $\mathcal{Z}$ value for each channel in different detectors and calculate the false alarm probability.
		The combined Bayes factors are found to be $\ln(\text{BF})_{\rm Fermi/GBM-NAI}=46.48$ and $\ln(\text{BF})_{\rm Swift/BAT}=18.98$, respectively.
		We assume that the prior odds $\pi_\text{lens}/\pi_\text{non-lens}$ is equal to $1/n$, where $n\sim 3,099~(1,297)$ is the number of GRBs detected by Fermi/GBM (Swift/BAT).
		Then we carry out model selection based on the posterior odds,}
	\begin{align}
		\mathcal{O}^\text{lens}_\text{non-lens} &= \frac{\mathcal{Z}_\text{lens}}{\mathcal{Z}_\text{non-lens}}
		\frac{\pi_\text{lens}}{\pi_\text{non-lens}} \\
		&= \frac{p_\text{lens} }{1 - p_\text{lens}}\\
		&= \frac{\text{BF}}{n},
	\end{align}
	{and the false alarm probability is
		$1-p_\text{lens}={1}/({1+\text{BF}/n})={1}/({1 + e^{\rm ln(BF)}/n})$, 
		which turns out to be 2.02 $\times$ $10^{-17}$ and 7.4 $\times$ $10^{-6}$ for Fermi/GBM-NAI and Swift/BAT data, respectively. Consequently, the lensing signal  is statistically significant. 
		
		With Eq.(\ref{mass_redshift}) and the posterior distributions of $\Delta t$ and $r$, the 90$\%$ credibility region of the redshifted mass $(1+z_{l})M_l$ are inferred to be $3.85^{+1.78}_{-0.94} \times 10^5 M_{\odot}$ for the Fermi/GBM-NAI data, $3.55^{+2.10}_{-0.99} \times 10^5 M_{\odot}$ for the Fermi/GBM-BGO data, $4.21^{+0.79}_{-0.68} \times 10^5 M_{\odot}$ for the Swift/BAT data and $5.75^{+1.93}_{-1.22} \times 10^5 M_{\odot}$ for the HXMT/HE data, respectively.
		The combination of the above redshifted mass distributions is $4.25^{+2.46}_{-1.36} \times 10^5 M_{\odot}$ (90\% credibility).}
	
	\subsection{Spectral analysis}
	We perform both time-integrated and time-resolved spectral analyses for GRB 200716C.
	The data of Fermi/GBM and Swift/BAT are used for joint spectral fittings.
	The pulse 1 and pulse 2 took place in the time intervals of [$T_{0}$, $T_{0}+0.60$ s] and [$T_{0}+2.00$ s, $T_{0}+2.60$ s], respectively.
	We further divide them into four or five slices to examine the temporal evolution of the spectra.
	An empirical smoothly-joined broken power-law function (the so-called ``Band'' function \citep{band1993batse}) and a CPL (cutoff power-law) function are adopted to fit the data.
	The Band function takes the form of  
	\begin{equation}
		N(E)=
		\begin{cases}
			A(\frac{E}{100\,{\rm keV}})^{\alpha}{\rm exp}{(-\frac{E}{E_0})} &\mbox{if $E<(\alpha-\beta)E_{0}$ }\\
			A[\frac{(\alpha-\beta)E_0}{100\,{\rm keV}}]^{(\alpha-\beta)}{\rm exp}{(\beta-\alpha)}(\frac{E}{100\,{\rm keV}})^{\beta} &\mbox{if $E > (\alpha-\beta)E_{0}$},
		\end{cases}
	\end{equation}
	where \emph{A} is the normalization constant, \emph{E} is the energy in unit of keV, $\alpha$ is the low-energy photon spectral index, $\beta$ is the high-energy photon spectral index, and \emph{E$_{0}$} is the break energy in the spectrum.
	The peak energy in the $\nu F_\nu$ spectrum is called $E_{p}$, which is equal to $(\alpha+2)E_{0}$.
	The CPL function is a power law with high energy exponential cutoff
	\begin{equation}
		{ N(E)=A(\frac{E}{100\,{\rm keV}})^{\alpha}{\rm exp}(-\frac{E}{E_c}) },
	\end{equation}
	where \emph{$\alpha$} is the power law photon spectral index, \emph{E$_{c}$} is the break energy in the spectrum,
	and the peak energy in the $\nu F_\nu$ spectrum is equal to $(\alpha+2)E_{c}$.
	
	\par
	The Bayesian information criterion \citep[BIC;][]{schwarz1978estimating} is adopted to evaluate the goodness of model fitting.
	The fitting results of different models in each time period are summarized in Table 2.
	Because of the low statistics of high-energy photons, $\beta$  can not be reliably constrained.
	By comparing the BIC values, we find that the CPL function is slightly preferred over the Band function.
	The temporal evolution of the spectral parameters are presented in the left middle (Band function) and lower (CPL funaction) panels  of Figure 1.
	The spectral parameters of the two pulses are similar. 
	The evolution of the time-resolved spectral properties of pulse 1 and pulse 2 are similar, too. Such facts are in support of the gravitational lensing model. 
	
	\par
	With the energy flux calculated by CPL function and the possible redshift ranging from $0.384$ \citep{2020GCN.28132....1D} to $5$, 
	we calculate the isotropic equivalent energy $E_{\gamma, {\rm iso}}$ with the cosmological parameters of \emph{H$_{0}$} = $\rm 69.6 ~kms^{-1}~Mpc^{-1}$, $\Omega_{\rm m}= 0.29$, and $\Omega_{\rm \Lambda}= 0.71$.
	In panel (a) of Figure 3, we compare GRB 200716C with some other GRBs with known redshift in the so-called Amati diagram \citep{amati2002intrinsic,zhang2009discerning,yang2020grb}.
	%\wyDel{The short GRBs follow the relation of log$E_{\rm p,z}= a+ b \log E_{\rm\gamma,iso}$ with the best-fitting parameters of $a=-12.82$ and $b=0.31$.}
	Clearly, GRB 200716C is well within the group of short GRBs, suggesting that the apparently long duration burst GRB 200716C may be a short event being lensed.
	
	%%%%%%%%%%%%%%%%%%%%%%%%%%%%%%%%%%%%%%%%%%%%%%%%%%%%%
	\subsection{Hardness test}
	We perform a simple but statistically powerful test, i.e., the cumulative hardness comparison in different energy bands \citep{mukherjee2021hardness}, for the presence of gravitational lensing in GRB 200716C. This is more intuitive than energy spectrum analysis by using a parameter ${\cal R}$ defined as
	\begin{equation}
		{\cal R}= \frac{P_{1,i}-B_i}{P_{2,i}-B_i},
	\end{equation}
	where $P_{1,i}$ and $P_{2,i}$ ($B_i$) represent the photon counts of the two pulses (background) over the duration in the $i$-th energy band.
	We use the data from Fermi/GBM (n0) and Swift/BAT for this test, and divide them into four bands as shown in the right panel of Figure 1. 
	To calculate the photon counts, the time intervals of the pulses are taken to be the same as that of the spectral analysis and the background $B$ is assumed to be a constant.
	The ratio errors include the Poisson noise imposed in the backgrounds as well as the pulses.
	Our results are presented in the panel (b) of Figure 3. 
	The resulting ${\cal R}$ in different channels are nearly constant for the Fermi/GBM and Swift/BAT data,
	consistent with the anticipation that the lensing images should have the same hardness.
	
	%%%%%%%%%%%%%%%%%%%%%%%%%%%%%%%%%%%%%%%%%%%%%%%%%%%%%
	\subsection{ Number density of lenses}
	
	With the eq.(29), eq.(33), eq.(34) and eq.(40) of \cite{paynter2021evidence}, we estimate the number density of the lenses with the Fermi/GBM data set characterized by the largest number of events ($N_{\rm GRB}=3099$). 
	%The symbols are the same as those defined in \cite{paynter2021evidence} unless specially mentioned.
	Since the time bin of our light curves is 16 ms, we set { the minimum time delay} of $\Delta t_{\rm min}=32~{\rm ms}$ (the results just change a little bit for $\Delta t_{\rm min}=16$ ms).
	In the calculation of the { maximum possible impact parameter} $y_{\rm max}$, the value of $\varphi_{\rm peak}/\varphi_0$ is the peak photon count rate for the second most brightly detector divided by the trigger threshold $kB^{1/2}$, where $k=4.5$ is the signal-to-noise ratio threshold set by the Fermi/GBM team \citep{meegan2009fermi}, and $B$ is the background under Poisson statistics.
	Based on the current data of Fermi/GBM, we { roughly} estimate the median of this ratio to be 1.6 and 5.7 for peak emission timescales of 16ms and 1024ms, respectively.
	We set $z_s$ = (5, 2, 0.5) and $z_l$ = (2.5, 1, 0.25) and take $M_l = 4.25^{+2.46}_{-1.36}\times 10^5$ M$_\odot$/$(1+z_l)$.  In the case of $\varphi_{peak}/\varphi_0 =1.6$,
	{ the range of the number density of lenses is inferred to be $n_l(z_s,z_l)=(0.18-12.20)\times10^{4}~{\rm Mpc^{-3}}$.}
	While for $\varphi_{peak}/\varphi_0$ =5.7, { we have a range of 
		$n_l(z_s,z_l)=(0.44-29.11)\times10^{3}~{\rm Mpc^{-3}}$}.		
	
	%%%%%%%%%%%%%%%%%%%%%%%%%%%%%%%%%%%%%%%%%%%%%%%%%%%%%
	
	\section{Conclusions and Discussion}
	In this work, we have analyzed the data of Fermi/GBM, Swift/BAT, and HXMT/HE to examine the possibility that GRB 200716C is actually a lensed short GRB.
	Our findings are the following:
	\begin{itemize}
		\item The light curves of the two pulses in all energy channels are well correlated with each other, as revealed in the 
		autocorrelation analysis as well as the Bayesian inference. 
		\item The temporal evolutions of the spectra of the two pulses are rather similar. In the Amati diagram, both pulses as well as the whole burst are well within the group of  short GRBs. 
		\item The measured flux ratios between the two pulses are nearly constant in all channels. 
	\end{itemize}
	
	Among the current GRB sample, such behaviors are very unusual.
	Intriguingly, all these facts can be straightforwardly understood in the gravitational lensing scenario. {Nevertheless, we can not completely rule out the possibility that this event is just a very special burst consisting of two intrinsically similar pulses.}
	If interpreted as a lensed GRB, the redshifted mass $(1+z_{l})M_l$ of the lens is about $4.25^{+2.46}_{-1.36} \times 10^5 M_{\odot}$ (90$\%$ credibility).
	For $z_{l}\leq 5$, 
	the foreground object would be an intermediate-mass black hole. 
	
	Different from the previous results with the sole BATSE data, our candidate is identified with the observations from Fermi/GBM, Swift/BAT, and HXMT/HE, which cover wider energy range and can be cross checked. We thus conclude that GRB 200716C is indeed a promising lensing event candidate. In this work, we focus on the specific event GRB 200716C. The analysis methods, however, can be directly applied to other sources. Our study of the large sample will be reported elsewhere.

	After the submission of this paper, the other work on GRB 200716C \citep{yang2021evidence}  also appeared in arXiv.
	In general, their analysis results are in agreement with ours except for a smaller lens mass. One reason may be that we have also included the BGO and HXMT data in the analyses, which will lead to different lens masses. These authors did not report the uncertainty range for the derived lens mass, rendering a more careful comparison difficult.

	\section*{Acknowledgments}
	We appreciate the anonymous referee and Dr. J. J. Wei for their helpful suggestions. We thank Dr. S. J. Lei for the kind help.
	We acknowledge the use of the public data from the {Fermi} archive, the Swift data archive, and the UK Swift Science Data Center. This work also made use of the data from the HXMT mission, a project funded by China National Space Administration (CNSA) and the Chinese Academy of Sciences (CAS).
	This work is supported by NSFC under grant No. 11921003.
	
	\bibliography{bibtex}
	\clearpage
	%%%%%%%%%%%%%%%%%%%%%%%%%%%%%%%%%%%%%%%%%%%%%%%%%%%%%
	\begin{figure}
		\centering
		\includegraphics[width=0.45\textwidth]{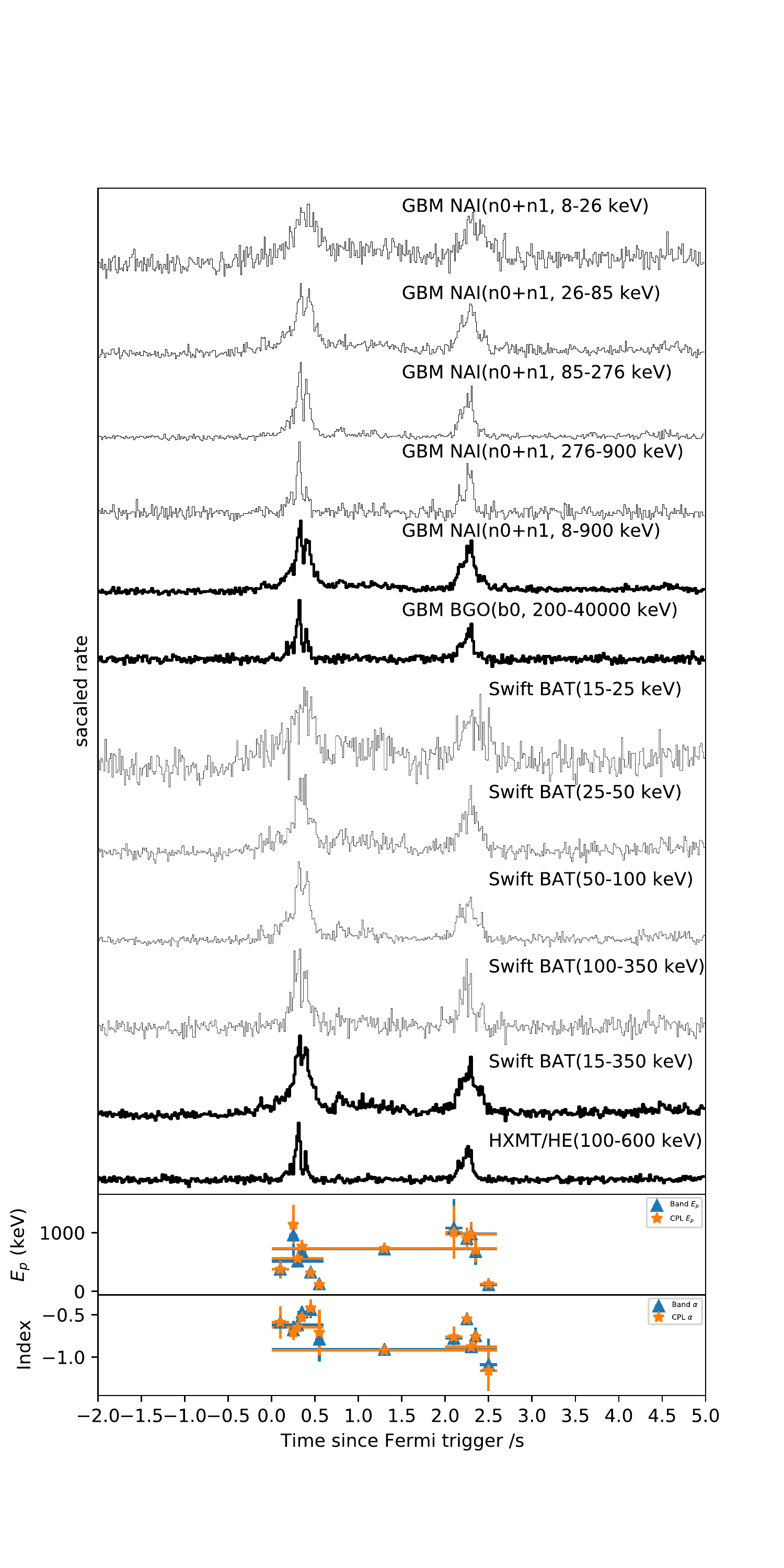}
		\includegraphics[width=0.45\textwidth]{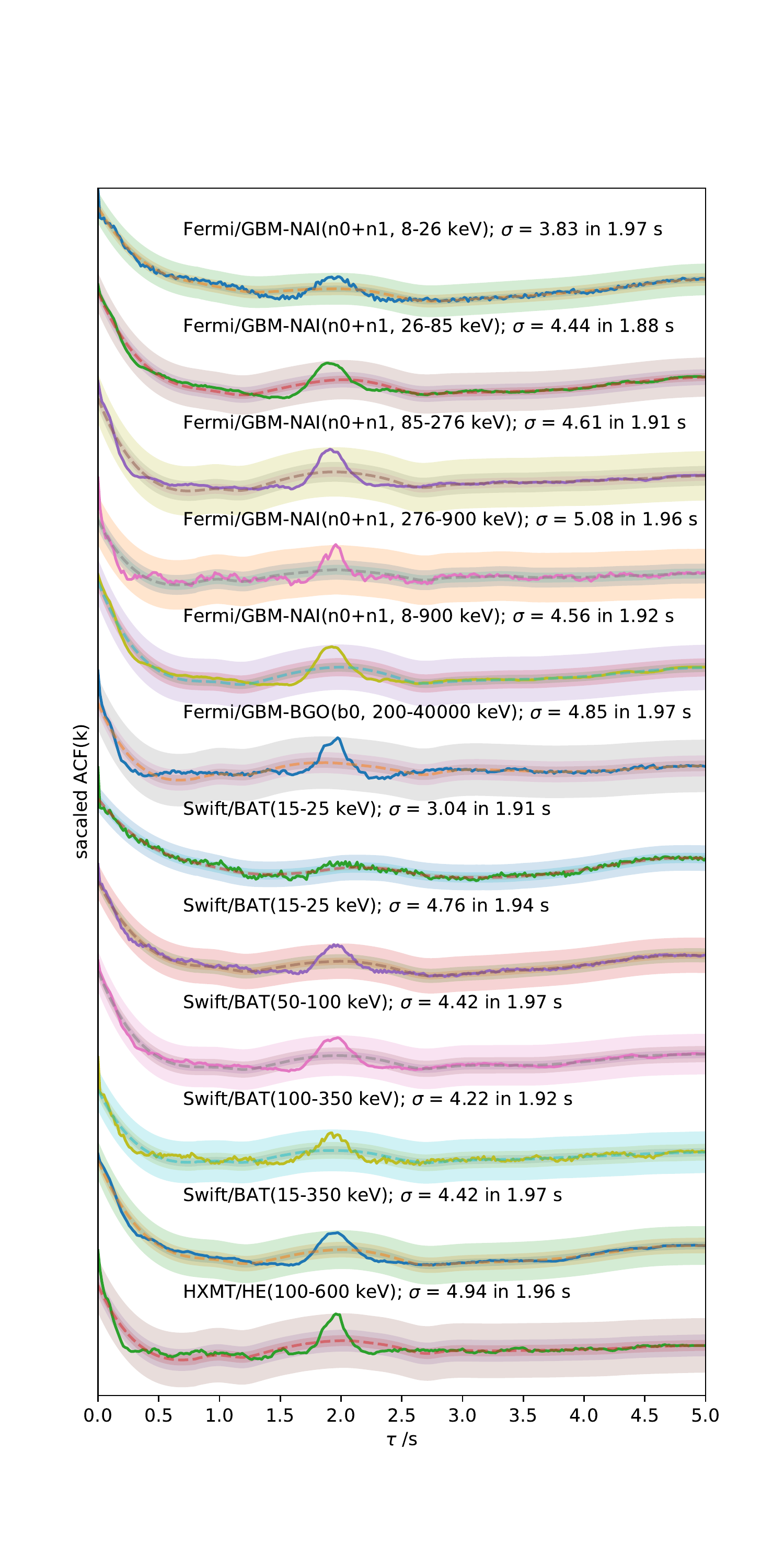}
		\caption{The left upper panel is the multi-energy band light curves of GRB 200716C observed by Fermi/GBM, Swift/BAT and HXMT/HE. The bin size of each band is 16 ms.
			The left middle and lower panels display the temporal evolution of spectral parameters $E_{\rm p}$ and $\alpha$. For  two pulses, the spectral parameters are similar.  
			The right panel is for the autocorrelation results of each energy band light curve of GRB 200716C.
			The dashed lines in different colors represent the fits to the light curves with a 3 order Savitzky-Golay smoothing filter.
			The shaded regions in colors  show the 1 $\sigma$, 3 $\sigma$, and 5 $\sigma$ containment bands of the Savitzky-Golay fit.
			The maximum significance and the corresponding time are also indicated in the plot.}
		\hfill
	\end{figure}
	%%%%%%%%%%%%%%%%%%%%%%%%%%%%%%%%%%%%%%%%%%%%%%%%%%%%%
	\begin{figure}
		\centering
		\includegraphics[width=0.7\textwidth]{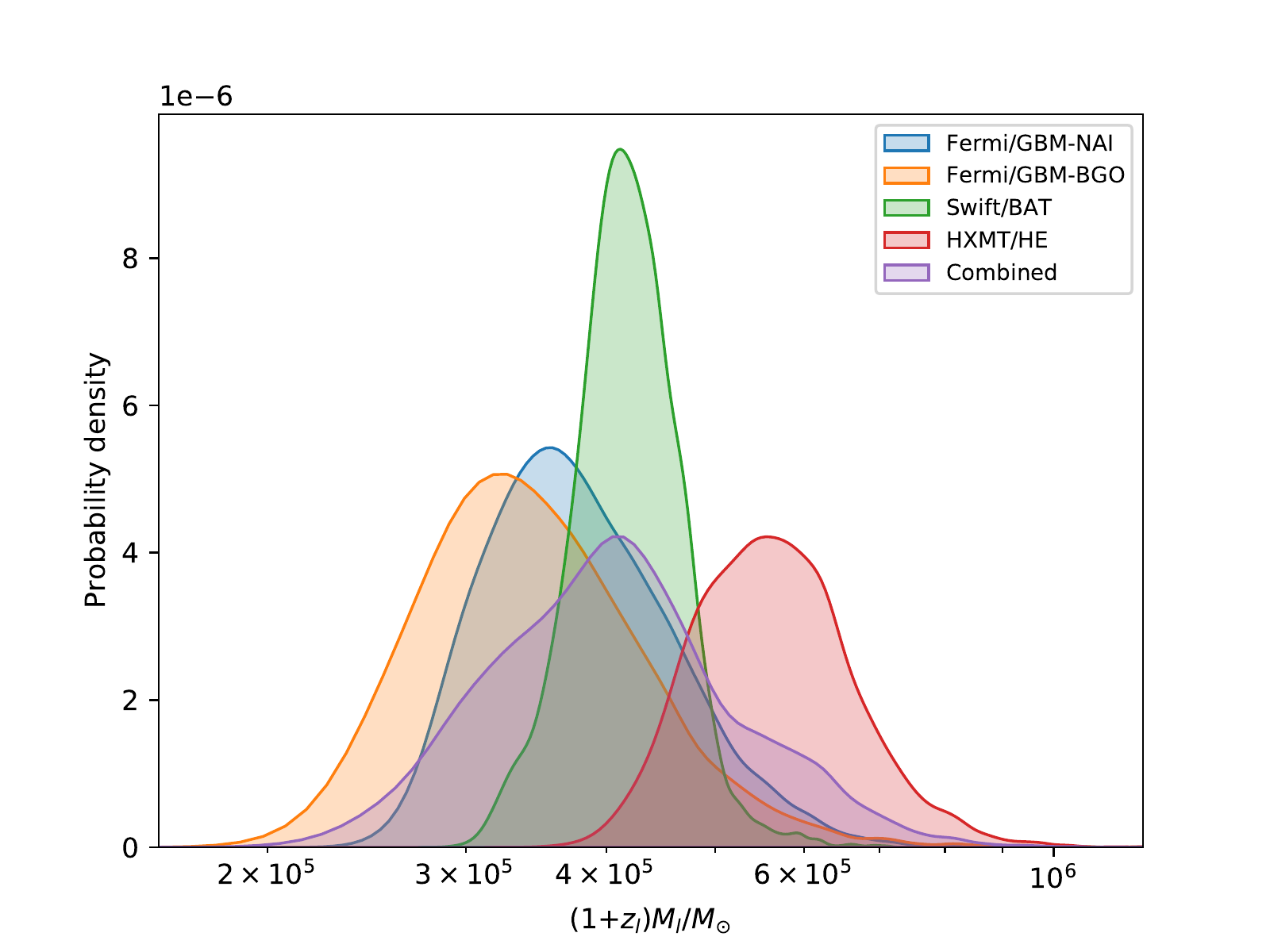}
		\caption{The redshifted lens masses inferred with different data sets. 
			The 90$\%$ credibility region of the redshifted mass $(1+z_{l})M_l$ are found to be $3.85^{+1.78}_{-0.94} \times 10^5 M_{\odot}$ for the Fermi/GBM-NAI data,
			$3.55^{+2.10}_{-0.99} \times 10^5 M_{\odot}$ for the Fermi/GBM-BGO data,
			$4.21^{+0.79}_{-0.68} \times 10^5 M_{\odot}$ for the Swift/BAT data and
			$5.75^{+1.93}_{-1.22} \times 10^5 M_{\odot}$ for the HXMT/HE data, respectively.
			The combination of the above redshifted mass distributions is $4.25^{+2.46}_{-1.36} \times 10^5 M_{\odot}$.}
		\hfill
	\end{figure}
	%%%%%%%%%%%%%%%%%%%%%%%%%%%%%%%%%%%%%%%%%%%%%%%%%%%%%
	\begin{figure}
		\centering
		\includegraphics[width=0.83\textwidth]{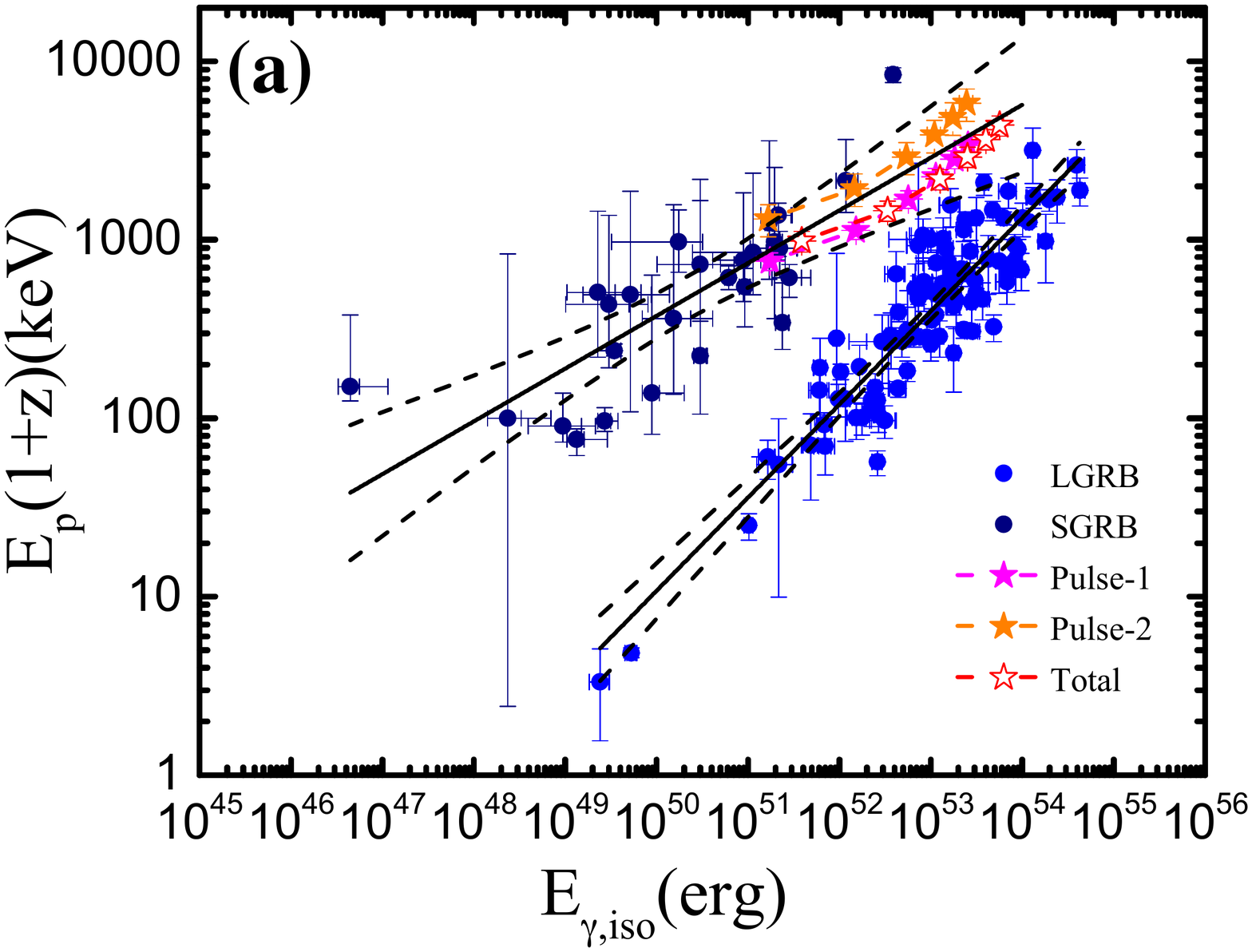}
		\includegraphics[width=0.63\textwidth]{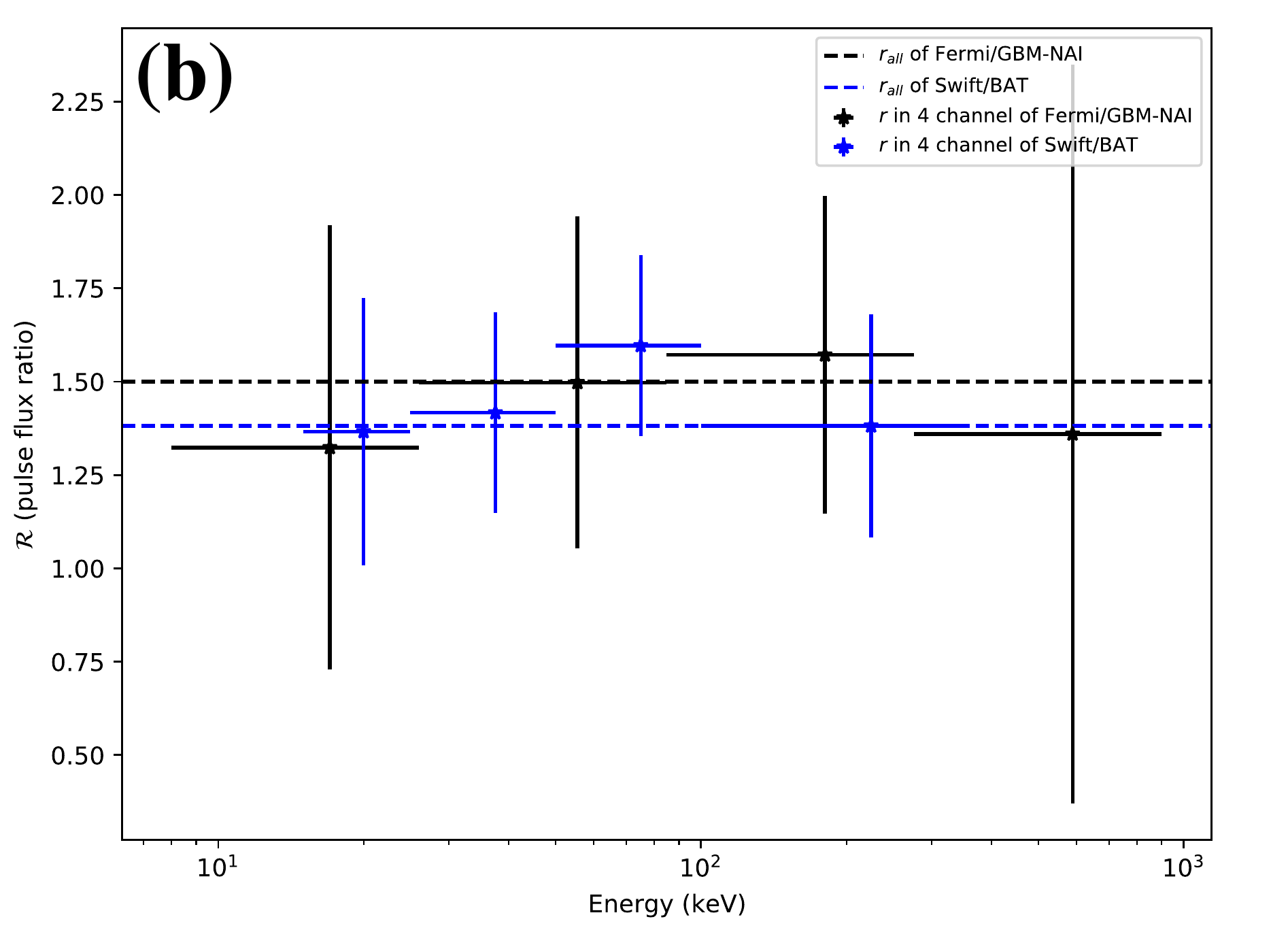}
		\caption{Two additional supports for the lensing scenario. (a) is the spectral peak energy ($E_{\rm p}$) and isotropic equivalent gamma-ray radiation energy ($E_{\rm\gamma,iso}$) correlation diagram.
			The pink and orange stars represent pulse 1 and pulse 2, respectively, and the hollow red star represents the whole burst.
			These stars from left to right are calculated for the assumed redshifts of 0.384, 1, 2, 3, 4, and 5, respectively.
			Both individual pulses and the whole burst are among the short GRB group rather than the long GRB group, which is consistent with the gravitational lensing model. 
			(b) is the cumulative hardness comparison results.
			For the Fermi/GBM and Swift/BAT data, the respective hardness ratios are consistent with being a constant, as anticipated in the lensing scenario. }
		\hfill
	\end{figure}
	%%%%%%%%%%%%%%%%%%%%%%%%%%%%%%%%%%%%%%%%%%%%%%%%%%%%%
	\clearpage
	\setlength{\tabcolsep}{2mm}{}
	\begin{deluxetable}{cccccccccccccccc}\tiny
		\tabletypesize{\tiny}
		\tablecaption{Bayesian inference results}
		\tablehead{ \colhead{Instrument and energy band}&\colhead{$\ln{\cal Z}_\text{lens}$} &\colhead{$\ln{\cal Z}_\text{lens,gaus}$} &\colhead{$\ln{\cal Z}_\text{non-lens}$} &\colhead{$\ln{\cal Z}_\text{non-lens,gaus}$} &\colhead{$\ln$(BF)}&\colhead{favourite model}&\colhead{}\\
		}
		\startdata
		Fermi/GBM-NAI\\
		{  8-26 keV} & -1327.67  & -1294.11 & -1319.56  & -1300.83  & 6.72  &$S_\text{lens,gaus}$ \\
		{  26-85 keV} & -1460.69  & -1394.89 & -1457.46  & -1400.78   & 5.89 &  $S_\text{lens,gaus}$\\
		{  85-276 keV} & -1447.15  & -1418.96 & -1446.00  & -1424.66   &  5.70 &$S_\text{lens,gaus}$\\
		{  276-900 keV} & -1021.75  & -1027.72 & -1031.51  & -1040.59  & 9.76  &$S_\text{lens}$\\
		{  Combined} & -5246.53  & -5120.38 & -5254.53  & -5166.86  & 46.48  &$S_\text{lens,gaus}$\\
		\hline
		Fermi/GBM-BGO\\
		%{  GBM NAI 8-900 keV} & -1930.31  & -1780.63 & -1922.06  & -1782.39   &   1.76 &$S_\text{lens}$\\
		{  200-40000 keV} & -1657.76  & -1661.03 & -1663.07  & -1666.65   &5.31   &$S_\text{lens}$\\
		\hline
		Swift/BAT\\
		{  5-25 keV} & -1365.42  & -1321.76 & -1360.21  & -1327.42   & 5.66 & $S_\text{lens,gaus}$ \\
		{  25-50 keV} & -1666.68  & -1503.67 & -1672.27  & -1502.47   &   -1.20 & $S_\text{non-lens,gaus}$\\
		{  50-100 keV} & -1740.42  & -1573.11 & -1735.68  & -1576.34   &  3.23 &  $S_\text{lens,gaus}$ \\
		{  100-350 keV} & -1419.00  & -1390.37 & -1419.97  & -1393.90   &  3.53 & $S_\text{lens,gaus}$ \\
		{  Combined} & -6182.75  & -5781.15 & -6188.13  & -5800.13  & 18.98  &$S_\text{lens,gaus}$\\
		\hline
		HXMT/HE\\
		%{  Swift BAT 15-350 keV} & -2629.45  & -2212.04 & -2622.10  & -2209.01   & -3.02  & $S_\text{non-lens+gaus}$\\
		{  100-600 keV} & -3024.81  & -3023.93 & -3024.84  & -3024.60   &  0.67 & $S_\text{lens,gaus}$ \\
		\enddata
		\tablecomments{The results of Bayesian inference of different models ($S_{\rm lens},S_{\rm lens,gaus},S_{\rm non-lens},S_{\rm non-lens,gaus}$).
			By comparing the $\ln{\cal Z}$ given by each observation instrument with Eq.(\ref{com_bayes}), we get the Bayes factor ($\ln({\rm BF})$) of the lensing model vs. non-lensing model,
			which are $46.48$, $5.31$, $18.98$, and $0.67$ for Fermi/GBM-NAI, Fermi/GBM-BGO, Swift/BAT, and HXMT/HE, respectively.
		}
	\end{deluxetable}
	%%%%%%%%%%%%%%%%%%%%%%%%%%%%%%%%%%%%%%%%%%%%%%%%%%%%%
	\setlength{\tabcolsep}{2mm}{}
	\begin{deluxetable}{cccccccccccccccc}\tiny
		\tabletypesize{\tiny}
		\tablecaption{Spectral analysis results for various slices}
		\tablehead{ \colhead{}&\colhead{} &\colhead{Band} &\colhead{} &\colhead{PG-stat/dof} &\colhead{BIC}&\colhead{}&\colhead{CPL}&\colhead{PG-stat/dof}&\colhead{BIC} &\colhead{}\\
			\colhead{Time interval} &  \colhead{$\alpha$}  &\colhead{$\beta$}&\colhead{$E_{\rm p}$} &\colhead{} &\colhead{} &\colhead{$\alpha$} &\colhead{$E_{\rm p}$} &\colhead{} &\colhead{}&\colhead{}\\
			\colhead{[s]} &\colhead{} &\colhead{}&\colhead{[keV]} &\colhead{} &\colhead{} &\colhead{} &\colhead{[keV]}}
		\startdata
		\object time-resolved spectra   \\
		\object Pulse 1   \\
		{  [0.00$- 0.20]$} & -0.59$\pm 0.19$  & -5.87$\pm200.83$ & 375.98$\pm155.44$  &337.97 /  409  &362.06  & -0.59$\pm 0.19$ & 376.13$\pm147.56$ &337.96 /  410 &356.03 \\
		{  [0.20$- 0.30]$} & -0.68$\pm 0.11$  & -2.13$\pm 0.25$ & 956.93$\pm356.94$  &318.48 /  409   &342.57  & -0.71$\pm 0.09$ & 1138.34$\pm334.47$ &323.53 /  410  &341.60 \\
		{  [0.30$- 0.40]$} & -0.47$\pm 0.07$  & -2.67$\pm 0.40$ & 674.49$\pm115.72$  &372.35 /  409   &396.44  & -0.53$\pm 0.06$ & 767.77$\pm114.52$ &373.50 /  410  &391.57 \\
		{  [0.40$- 0.50]$} & -0.44$\pm 0.10$  & -5.99$\pm83.10$ & 325.72$\pm55.49$  &342.33 /  409  &366.42  & -0.42$\pm 0.10$ & 313.12$\pm49.83$ &342.12 /  410 &360.19 \\
		{  [0.50$- 0.60]$} & -0.79$\pm 0.26$  & -6.00$\pm402.22$ & 126.22$\pm62.25$  &272.38 /  409  &296.47  & -0.71$\pm 0.27$ & 112.96$\pm49.34$ &272.00 /  410  &290.07 \\
		\object Pulse 2 \\
		{  [2.00$- 2.20]$} & -0.78$\pm 0.12$  & -5.28$\pm44.61$ & 1079.07$\pm497.45$  &346.11 /  409   &370.21  & -0.76$\pm 0.12$ & 1007.62$\pm446.67$ &346.05 /  410  &364.13 \\
		{  [2.20$- 2.30]$} & -0.55$\pm 0.08$  & -5.90$\pm77.74$ & 901.38$\pm183.47$  &354.93 /  409   &379.02  & -0.55$\pm 0.08$ & 898.67$\pm180.31$ &354.91 /  410  &372.98 \\
		{  [2.30$- 2.40]$} & -0.75$\pm 0.10$  & -2.81$\pm 1.29$ & 677.90$\pm229.22$  &320.51 /  409   &344.60  & -0.76$\pm 0.09$ & 703.42$\pm207.07$ &321.00 /  410  &339.07 \\
		{  [2.40$- 2.60]$} & -1.09$\pm 0.31$  & -2.61$\pm 1.50$ & 112.11$\pm94.75$  &393.96 /  409   &418.05  & -1.16$\pm 0.23$ & 127.51$\pm77.91$ &393.87 /  410  &411.94 \\
		\hline
		\object time-integrated spectra   \\
		\object Pulse 1   \\
		{  [0.00$- 0.60]$} & -0.62$\pm 0.05$  & -2.51$\pm 0.30$ & 514.50$\pm69.87$  &382.44 /  409   &406.53  & -0.65$\pm 0.05$ & 564.17$\pm64.09$ &384.74 /  410  &402.81 \\
		\object Pulse 2   \\
		{  [2.00$- 2.60]$} & -0.88$\pm 0.05$  & -4.42$\pm 8.17$ & 976.51$\pm204.22$  &354.95 /  409   &379.05  & -0.88$\pm 0.05$ & 969.44$\pm200.81$ &354.97 /  410  &373.04 \\
		\object Total   \\
		{  [0.00$- 2.60]$} & -0.91$\pm 0.03$  & -3.33$\pm 1.39$ & 720.15$\pm99.16$  &349.14 /  409   &373.23  & -0.92$\pm 0.03$ & 730.34$\pm94.94$ &349.33 /  410  &367.40 \\
		\enddata
		\tablecomments{This table summarizes the results of the energy spectrum analysis for each time interval, including the parameters of the Band function ($\alpha, \beta, E_p$) and the CPL function ($\alpha, E_p$), and the value that characterizes the goodness of fit (BIC).
			For the detailed meaning of each parameter, see Section 3.2.
			In addition, in each time slice, the goodness of fit of the CPL function is better than that of the Band function. 
			That is why the results of the CPL function are used to calculate the energy fluence.
		}
	\end{deluxetable}
	%%%%%%%%%%%%%%%%%%%%%%%%%%%%%%%%%%%%%%%%%%%%%%%%%%%%%
\end{document}